# Chromodynamic Weibel instabilities in relativistic nuclear collisions

Jørgen Randrup[1] and Stanisław Mrówczyński[2]

[1]*Nuclear Science Division, Lawrence Berkeley National Laboratory, Berkeley, California 94720, U.S.A.*
[2]*Sołtan Institute for Nuclear Studies, ul. Hoża 69, PL-00-681 Warsaw, Poland, and
Institute of Physics, Świętokrzyska Academy, ul. Świętokrzyska 15, PL-25-406 Kielce, Poland*

(Dated: March 10, 2003; revised June 30, 2003)

Employing a previously derived formulation, and extending the treatment from purely transverse modes to wave vectors having a longitudinal component, we discuss the prospects for the occurrence of Weibel-type color-current filamentation in high-energy nuclear collisions. Numerical solutions of the dispersion equation for a number of scenarios relevant to RHIC and LHC suggest that modes with (predominantly transverse) wave numbers of several hundred MeV may become moderately agitated during the early collision stage. The emergence of filamentation helps to speed up the equilibration of the parton plasma and it may lead to non-statistical azimuthal patterns in the hadron final state.

PACS numbers: 12.38.-t 24.85.+p 25.75.-q, 52.35.-g

## I. INTRODUCTION

In the exploration of high-energy nuclear collisions [1], the possible occurrence of collective phenomena is of particular interest. The present note reports on studies of color filamentation during the early collision stage.

Within the framework of electrodynamics, Weibel [2] pointed out that self-excited transverse modes exist in plasmas with anisotropic momentum distributions and he derived their rate of growth from the linear response of the collisionless Boltzmann transport equation (also known as the Vlasov equation). This treatment was later adapted to counterstreaming fluids of nucleons [3, 4] or partons [5–9] and it was found that also these systems possess the Weibel filamentation instability. Although idealized counterstreaming may not be realized, it is expected that a significant degree of local momentum-space anisotropy will occur at the early stages of relativistic nuclear collisions at RHIC or LHC and investigations have been carried out of the associated Weibel filamentation modes in chromodynamic plasmas [10–12].

The present work employs the approach formulated in these latter investigations in an attempt to achieve a more complete and quantitative understanding of the Weibel filamentation phenomenon in high-energy nuclear collisions. After recalling the most important developments made in Refs. [10–13], we first discuss the general features of the filamentation phenomenon and considering, for the first time, modes that are not simply transversally aligned. For plasmas relevant to RHIC or LHC, we present numerical results for the resulting growth rates. Furthermore, for suitable idealized dynamical scenarios, we extract the collective amplification coefficients and elucidate the importance of the rapid longitudinal expansion as well as the equilibration caused by collisions among the partons. We conclude by discussing the possible dynamical implications of these collective modes.

## II. FORMAL FRAMEWORK

The present study is based on the developments made previously in Refs. [10–13] and we recall here the most relevant elements of those. The treatment is made within the semi-classical transport framework in which the partons are described by their phase-space densities. An early review of quark-gluon transport theory was given by Elze and Heinz [14], while a very recent review may be found in Ref. [15].

In the present problem, we perturb a plasma of quarks, antiquarks, and gluons whose phase-space densities are uniform in space and stationary in time. Furthermore, they are distributed equally among the various color channels so the corresponding background densities are color singlets. For simplicity, the partons are assumed to be massless so their energies are $E_p = |\boldsymbol{p}|$.

There are, in principle, separate phase-space densities for each quark flavor and each spin component, but since they all contribute additively we may consider just one generic quark density with a four-fold flavor-spin degeneracy, $Q(x, \boldsymbol{p})$, where $x \equiv (t, \boldsymbol{r})$ denotes the four-position and $\boldsymbol{p}$ is the quark three-momentum. Similarly, we consider just one common phase-space density for the antiquarks, $\bar{Q}(x, \boldsymbol{p})$.

In addition to being $4 \times 4$ Dirac tensors, the phase-space densities $Q$ and $\bar{Q}$ are $N \times N$ color matrices in the SU($N$) gauge group that have singlet (colorless) and multiplet (colored) parts,

$$Q(x, \boldsymbol{p}) = Q_0(\boldsymbol{p}) + \delta Q(x, \boldsymbol{p}) , \qquad (1)$$
$$\bar{Q}(x, \boldsymbol{p}) = \bar{Q}_0(\boldsymbol{p}) + \delta \bar{Q}(x, \boldsymbol{p}) . \qquad (2)$$

The singlet parts represent the $q$ and $\bar{q}$ background phase-space densities, $Q_0 = If_q$ and $\bar{Q}_0 = If_{\bar{q}}$ (where $I$ is the $N \times N$ unit matrix in color space), with the phase-space number densities for each color being

$$f_q(\boldsymbol{p}) = \frac{1}{N}\text{Tr}[Q(x,\boldsymbol{p})] , \; f_{\bar{q}}(\boldsymbol{p}) = \frac{1}{N}\text{Tr}[\bar{Q}(x,\boldsymbol{p})] . \qquad (3)$$

The induced disturbances, $\delta Q$ and $\delta \bar{Q}$, which represent deviations from the color neutrality, are assumed to be much smaller than the colorless background terms. They may be expanded on the $N^2-1$ SU($N$) group generators $\{t_a\}$, which satisfy $\text{Tr}[t_a t_b] = \frac{1}{2}\delta_{ab}$ (the trace is with respect to the color indices), and the individual color components, $Q_a$ and $\bar{Q}_a$, can be extracted by projection,

$$Q_a(x,\boldsymbol{p}) = 2\text{Tr}[t_a Q(x,\boldsymbol{p})] \ , \ \bar{Q}_a(x,\boldsymbol{p}) = 2\text{Tr}[t_a \bar{Q}(x,\boldsymbol{p})] \ . \quad (4)$$

Because the group generators are traceless, only the colorless part of $Q$, i.e. $Q_0$, contributes to $f_q$, while only the colored part $\delta Q$ contributes to $Q_a$.

The gluon phase-space density $\mathcal{G}(\boldsymbol{r},\boldsymbol{p})$ is an $(N^2-1) \times (N^2-1)$ matrix in the adjoint representation which is spanned by the $N^2-1$ matrices $\{T^a\}$, whose elements are the SU($N$) structure constants, $(T^a)_{bc} = -if_{abc}$. The identity $f_{abc}f_{dbc} = N\delta_{ad}$ implies that $\text{Tr}[T^a T^b] = N\delta^{ab}$. We may thus write

$$\mathcal{G}(x,\boldsymbol{p}) = \mathcal{G}_0(\boldsymbol{p}) + \delta\mathcal{G}(x,\boldsymbol{p}) = f_g(\boldsymbol{p})\mathcal{I} + T_a \mathcal{G}_a(x,\boldsymbol{p}) \ , \quad (5)$$

where the individual elements are

$$\mathcal{G}_{ab}(x,\boldsymbol{p}) = f_g(\boldsymbol{p})\delta_{ab} - if_{abc}\mathcal{G}_c(x,\boldsymbol{p}) \ . \quad (6)$$

The color singlet part, $\mathcal{G}_0$, is the uniform and stationary gluon background distribution and $f_g(\boldsymbol{p})$ is the associated phase-space number density in each of the $N^2-1$ different color channels labeled by $a$. The various color components can be extracted by trace operations,

$$f_g(\boldsymbol{p}) = \frac{1}{N^2-1}\text{Tr}[\mathcal{G}(x,\boldsymbol{p})] = \frac{1}{N^2-1}\mathcal{G}_{aa}(x,\boldsymbol{p}) \ , \quad (7)$$

$$\mathcal{G}_a(x,\boldsymbol{p}) = \frac{1}{N}\text{Tr}[T_a \mathcal{G}(x,\boldsymbol{p})] = \frac{i}{N}f_{abc}\mathcal{G}_{bc}(\boldsymbol{r},\boldsymbol{p}) \ . \quad (8)$$

The disturbances in the parton phase-space densities contribute additively to the induced current densities $j_a$,

$$j_a^\mu(\boldsymbol{r},t) = \tfrac{1}{2}g\int \frac{d^3\boldsymbol{p}}{(2\pi)^3}\frac{p^\mu}{E_{\boldsymbol{p}}}\left[Q_a(\boldsymbol{r},\boldsymbol{p},t) - \bar{Q}_a(\boldsymbol{r},\boldsymbol{p},t) \right.$$
$$\left. + 2if_{abc}\mathcal{G}_{cb}(\boldsymbol{r},\boldsymbol{p},t)\right] \ , \quad (9)$$

where $g$ is the QCD coupling constant, $g^2 = 4\pi\hbar c\alpha_s$.

In the collisionless idealization, the parton phase-space densities satisfy Vlasov-type transport equations,

$$p^\mu D_\mu Q(x,\boldsymbol{p}) + \frac{g}{2}p^\mu\left\{F_{\mu\nu}(x), \frac{\partial Q(x,\boldsymbol{p})}{\partial p_\nu}\right\} = 0 , (10)$$

$$p^\mu D_\mu \bar{Q}(x,\boldsymbol{p}) - \frac{g}{2}p^\mu\left\{F_{\mu\nu}(x), \frac{\partial \bar{Q}(x,\boldsymbol{p})}{\partial p_\nu}\right\} = 0 , (11)$$

$$p^\mu \mathcal{D}_\mu \mathcal{G}(x,\boldsymbol{p}) + \frac{g}{2}p^\mu\left\{\mathcal{F}_{\mu\nu}(x), \frac{\partial \mathcal{G}(x,\boldsymbol{p})}{\partial p_\nu}\right\} = 0 , (12)$$

where $\{\cdot,\cdot\}$ denotes the anticommutator. The covariant derivatives $D_\mu$ and $\mathcal{D}_\mu$ act as follows,

$$D_\mu = I\partial_\mu - ig[A_\mu(x),\cdot] \ , \ \mathcal{D}_\mu = \mathcal{I}\partial_\mu - ig[\mathcal{A}_\mu(x),\cdot] \ , \quad (13)$$

with $A_\mu$ and $\mathcal{A}_\mu$ being the potentials in the fundamental and adjoint representations, respectively,

$$A^\mu(x) = A_a^\mu(x)\tau_a \ , \ \mathcal{A}_{ab}^\mu(x) = -if_{abc}A_c^\mu(x) \ . \quad (14)$$

The stress tensor in the fundamental representation is $F_{\mu\nu} = \partial_\mu A_\nu - \partial_\nu A_\mu - ig[A_\mu, A_\nu]$, while $\mathcal{F}_{\mu\nu}$ denotes the field strength tensor in the adjoint representation.

With the phase-space densities of the form (1,2,5), the transport equations, which are linearized with respect to $\delta Q$, $\delta\bar{Q}$ and $\delta\mathcal{G}$, can be explicitly solved in terms of the free Green's functions. Substituting these solutions into the expression (9) for the current, one finds the Fourier transform of the induced current,

$$j_a^\mu(k) = \Pi^{\mu\nu}(k)\,A_{a;\nu}(k) \ , \quad (15)$$

where $k \equiv (\omega,\boldsymbol{k})$. The polarization tensor $\Pi^{\mu\nu}(k)$ does not carry a color index, since the potential in a given color channel, $A_a^\mu$, can induce color currents in only that channel $a$. It is given by

$$\Pi^{\mu\nu}(k) = g^2\int\frac{d^3\boldsymbol{p}}{(2\pi)^3}\frac{p^\mu}{E_{\boldsymbol{p}}}\left[g^{\nu\lambda} - \frac{p^\nu k^\lambda}{p^\sigma k_\sigma + i\epsilon}\right]\frac{\partial f_{\text{eff}}(\boldsymbol{p})}{\partial p^\lambda} \ , \quad (16)$$

with the *effective* background phase-space density being

$$f_{\text{eff}}(\boldsymbol{p}) \equiv \tfrac{1}{2}f_q(\boldsymbol{p}) + \tfrac{1}{2}f_{\bar{q}}(\boldsymbol{p}) + Nf_g(\boldsymbol{p}) \ . \quad (17)$$

The result (15) is similar to the electromagnetic case, the only difference being the replacement of the background phase-space density of charge carriers, $f_{e+\bar{e}}(\boldsymbol{p})$, by the *effective* background phase-space density contributing to each of the $N^2-1$ individual color channels $a$. In order to simplify the analysis, we assume that all the parton species (quarks, antiquarks, and gluons) have the same momentum profile, $\phi(\boldsymbol{p})$, with $\int d^3\boldsymbol{p}\,\phi(\boldsymbol{p}) = 1$. So, $f_{\text{eff}}(\boldsymbol{p}) \approx \rho_{\text{eff}}\phi(\boldsymbol{p})$ where the effective parton density is $\rho_{\text{eff}} = \tfrac{1}{2}(\rho_q + \rho_{\bar{q}}) + N\rho_g$.

Insertion of the expression (15) for the induced current into the field equation of motion, $D_\mu F^{\mu\nu}(x) = j^\nu(x)$, then yields an algebraic equation for $A_a^\mu(k)$,

$$\left[k^2 g^{\mu\nu} - k^\mu k^\nu - \Pi^{\mu\nu}(k)\right]A_{a;\nu}(k) = 0 \ , \quad (18)$$

which in turn leads to the dispersion equation for the collective modes,

$$\det\left[k^2 g^{\mu\nu} - k^\mu k^\nu - \Pi^{\mu\nu}(k)\right] = 0 \ . \quad (19)$$

However, due to the transversality of the polarization tensor, $k_\mu \Pi^{\mu\nu}(k) = 0$, the above equation involving a $4 \times 4$ determinant can be reduced to an equation that involves only a $3\times 3$ determinant. This simplification can be most easily understood in the Coulomb gauge, where the chromoelectric three-vector is given by $\boldsymbol{E}_a = \partial_t \boldsymbol{A}_a$. Then Eq. (18) for $A_a^\mu$ is immediately transformed into an equation for $\boldsymbol{E}_a$ and the dispersion equation for the collective modes then readily follows,

$$\det[\mathbf{k}^2\delta^{ij} - k^i k^j - \omega^2 \epsilon^{ij}(k)] = 0 \ , \quad (20)$$

where the $3 \times 3$ color permittivity tensor $\boldsymbol{\epsilon}$ has the elements $\epsilon^{ij}(k) = \delta^{ij} - \omega^{-2}\Pi^{ij}(k)$. Although the dispersion equation (20) has been derived here in the Coulomb gauge, it is gauge independent since the position of a pole in the gluon propagator is the same for all gauges, even though the gluon propagator itself is a gauge dependent quantity (see Ref. [15], for example).

In the subsequent presentation, we study Eq. (20) with the polarization tensor derived in the collisionless limit, in which the treatment is simplest. A general formal justification for adopting this scenario as a first approximation lies in the fact that the frequency of the plasma waves is of order $g$, whereas that of the binary collisions is of order $g^2$ or $g^4$, depending on the momentum transfer. Thus, for sufficiently small coupling constants $g$, the collective modes have a shorter characterisitc time than the collisions so, consequently, the collisions do not influence the waves for short periods of time. However, in the present case, where the focus is on unstable modes, this argument may not hold, since the characteristic growth rates tend to be significantly slower than the frequencies of the stable modes. Furthermore, the coupling constant is actually not that small and the parton density is initially relatively high. Therefore, the dispersion equation (20) derived within the collisionless idealization should be regarded merely as a starting point for obtaining a rough idea of the quantitative importance of the phenomenon. If it appears warranted, more refined treatments would be called for.

## III. GENERAL FEATURES

In the present study, we consider momentum profiles $\phi(\boldsymbol{p})$ that are invariant with respect to rotations around the $z$ axis as well as reflections in the $xy$ plane (it then follows that $\phi(-\boldsymbol{p}) = \phi(\boldsymbol{p})$). Then the profile function depends only on the magnitude of the momentum component along the symmetry axis, $p_\parallel = |p_z|$, and the magnitude of the transverse momentum, $p_\perp = (p_x^2 + p_y^2)^{1/2}$. Due to this symmetry, we may generally assume that the wave vector is of the form $\mathbf{k} = (k_x, 0, k_z)$, with $k_x = k_\perp > 0$ and $k_z = k_\parallel \geq 0$. Finally, since we are seeking modes for which the frequency is purely imaginary, we write $\omega = i\gamma$ where $\gamma$ is real.

The elements of the polarization tensor $\boldsymbol{\Pi}(k)$ are

$$\Pi^{ij}(k) = g^2 \rho_{\text{eff}} \langle -v^i \frac{\partial}{\partial p^j} + \frac{v^i v^j}{(\mathbf{k} \cdot \boldsymbol{v})^2 + \gamma^2} \mathbf{k} \cdot \boldsymbol{v} \, k^l \frac{\partial}{\partial p^l} \rangle \ , \quad (21)$$

where $\langle \cdot \rangle$ denotes the average over the profile function $\phi(\boldsymbol{p})$. Since $\Pi_{xy}$ and $\Pi_{yz}$ vanish in the present case, the determinant (20) factorizes into a $y$ part, $\mathbf{k}^2 + \gamma^2 + \Pi_{yy}$, which has no real roots, and an $xz$ part presenting the dispersion equation of interest, $D_{\mathbf{k}}(\gamma) = 0$, where

$$D_{\mathbf{k}} \equiv (k_z^2 + \gamma^2 + \Pi_{xx})(k_x^2 + \gamma^2 + \Pi_{zz}) - (k_x k_z - \Pi_{xz})^2. \quad (22)$$

Since $D_{\mathbf{k}}$ depends on $\gamma$ only through its square, the roots come in opposite pairs, $\omega_{\mathbf{k}} = \pm i\gamma_{\mathbf{k}}$. Furthermore, it can be seen that $D_{\mathbf{k}}$ vanishes for $\gamma = 0$ since we then have

$$\begin{pmatrix} \Pi_{xx} & \Pi_{xz} \\ \Pi_{zx} & \Pi_{zz} \end{pmatrix} = \begin{pmatrix} k_z \langle v_x [\cdots] \rangle & -k_x \langle v_x [\cdots] \rangle \\ k_z \langle v_z [\cdots] \rangle & -k_x \langle v_z [\cdots] \rangle \end{pmatrix}, \quad (23)$$

where $[\cdots] \equiv [g^2 \rho_{\text{eff}}(-v_z \partial/\partial p_x + v_x \partial/\partial p_z)/\mathbf{k} \cdot \boldsymbol{v}]$. The $\gamma$ dependence of $\boldsymbol{\Pi}$ notwithstanding, $D_{\mathbf{k}}(\gamma)$ appears qualitatively as a fouth-order polynomial in $\gamma$, with $D_{\mathbf{k}} \sim \gamma^4$ for large $|\gamma|$. Consequently, in addition to the double root at zero, there is at most one other pair of real roots, $\pm\gamma_{\mathbf{k}}$. Thus, for any given wave vector $\mathbf{k}$, the dispersion equation (22) has at most one positive root, $\gamma_{\mathbf{k}}$.

When the wave vector is perpendicular to the axis of symmetry, $\mathbf{k} = (k_\perp, 0, 0)$, the polarization tensor is fully diagonal, $\Pi_{xz} = 0$. Furthermore, $\Pi_{xx} > 0$, so the dispersion equation (22) reduces to

$$0 = k_\perp^2 + \gamma^2 + \Pi_{zz} \ , \text{ or } k_\perp^2 = \omega^2 \epsilon_{zz} \ . \quad (24)$$

This equation has a positive root as long as $k_\perp < k_{\max}$, the maximum being determined by $k_{\max}^2 = -\Pi_{zz}(\gamma = 0)$,

$$k_{\max}^2 = 2g^2 \rho_{\text{eff}} \langle \frac{p_z^2}{E} \left( \frac{\partial}{\partial p_z^2} - \frac{\partial}{\partial p_x^2} \right) \rangle \ . \quad (25)$$

This relation shows that there are Weibel filamentation modes as long as the momentum profile falls off "more rapidly" in the transverse direction than longitudinally.

It is readily seen that the polarization tensor remains unaffected under a rescaling of its four-vector argument $k = (\omega, \mathbf{k})$, $\boldsymbol{\Pi}(ak) = \boldsymbol{\Pi}(k)$. Thus, since $\boldsymbol{\Pi}(k)$ is proportional to $g^2 \rho_{\text{eff}}$ the entire dispersion equation remains unchanged if at the same time that quantity is multiplied by $a^2$. Consequently, *if* for any given value of $g^2 \rho_{\text{eff}}$, the mode with the wave vector $\mathbf{k}$ has the growth rate $\gamma_{\mathbf{k}}$, *then* in a plasma where the value of $g^2 \rho_{\text{eff}}$ is changed by the factor $a^2$, the mode with the wave vector $a\mathbf{k}$ has the growth rate $a\gamma_{\mathbf{k}}$. This general scaling property is very useful, as it makes it possible to significantly extend the utility of results obtained for specific scenarios.

As $k_\perp$ is varied from zero to $k_{\max}$ (keeping $k_\parallel = 0$), the growth rate $\gamma_{\mathbf{k}}$ starts out linearly from zero, has a broad maximum, and then decreases towards zero again. The initial slope, $(d\gamma/dk_\perp)_0$, is determined by

$$\langle v_z \frac{\partial}{\partial p_z} \rangle = \langle \frac{v_z^2}{v_x^2 + (d\gamma/dk_\perp)_0^2} v_x \frac{\partial}{\partial p_x} \rangle \ , \quad (26)$$

and it increases with the degree of anisotropy.

For any given value of $k_\perp$ for which the transverse dispersion equation (24) has a solution for $\mathbf{k} = (k_\perp, 0, 0)$ (*i.e.* for $k_\perp < k_{\max}$), there exists a range of values $k_\parallel < k_\parallel^{\max}(k_\perp)$ for which the general dispersion equation (22) has solutions for $\mathbf{k} = (k_\perp, 0, k_\parallel)$. [For $k_\parallel = 0$ the determinant $D_{\mathbf{k}}$, considered as a function of $\gamma^2$, starts out from zero and exhibits a negative minimum before starting its steady growth towards $\sim \gamma^2$. The value of $\gamma$ for which $D_{\mathbf{k}}$ becomes positive is the associated growth rate $\gamma_{\mathbf{k}}$. As $k_\parallel$ is increased, the minimum moves inwards

(as does the crossing point $\gamma_\mathbf{k}$) and it eventually merges with the maximum at zero. We then have $\partial^2 D_\mathbf{k}/\partial\gamma^2 = 0$ and $k_\parallel$ has attained its maximum value $k_\parallel^{\max}(k_\perp)$.]

As the direction of the wave vector $\mathbf{k}$ is moved out of the transverse $(xy)$ plane, the direction of the electric field vector $\mathbf{E}$ starts to deviate from the $z$ axis, as does the direction of the induced current $\mathbf{j}$. In the non-Abelian plasma, the specific behavior is gauge dependent and we shall illustrate it in the Coulomb gauge (Sect. VI).

## IV. CALCULATIONAL SCENARIOS

In order to have a concrete framework for the presentation of the numerical results, we shall adopt the following standard values for the RHIC and LHC scenarios [16, 17],

$$\text{RHIC: } \alpha_s = 0.3 \,, \; \rho_{\text{eff}}(t_0 = 0.4 \text{ fm}/c) = 6 \text{ fm}^{-3}, \quad (27)$$
$$\text{LHC: } \alpha_s = 0.1 \,, \; \rho_{\text{eff}}(t_0 = 0.3 \text{ fm}/c) = 50 \text{ fm}^{-3}. \quad (28)$$

Obviously, they represent very rough estimates only and we have tried to err on the conservative side. Larger values for either the coupling constant or the parton density would only enhance the filamentation effect and the simple scaling properties of the dispersion equation (22) makes it relatively easy to determine what would result if these input values were changed.

In our studies, we shall employ two different analytical parametrizations of the momentum profile function $\phi(\mathbf{p})$. The first form is a simple Gaussian,

$$\phi^{\text{Gauss}}(\mathbf{p}) = \frac{1}{\pi\sigma_\perp^2}\frac{1}{\sqrt{2\pi}\sigma_z} e^{-\frac{p_x^2+p_y^2}{\sigma_\perp^2} - \frac{p_z^2}{2\sigma_\parallel^2}}, \quad (29)$$

where $\sigma_\perp^2 = \langle p_x^2 + p_y^2 \rangle = \sigma_x^2 + \sigma_y^2$ and $\sigma_\parallel^2 = \langle p_z^2 \rangle = \sigma_z^2$. Since the results will depend only on the anisotropy $\sigma_z/\sigma_x$, we shall adopt a fixed value for the transverse variance, $\sigma_\perp = 300$ MeV$/c$, and then vary the longitudinal width $\sigma_\parallel$. We note that $\langle \partial/\partial p_x^2 \rangle = \langle \partial/\partial p_y^2 \rangle = -\sigma_\perp^{-2}$ while $\langle \partial/\partial p_z^2 \rangle = -\frac{1}{2}\sigma_\parallel^{-2}$ for the Gaussian profile.

As an alternative, we employ a pQCD-motivated profile, which has a polynomial transverse falloff,

$$\phi^{\text{pQCD}}(\mathbf{p}) = \frac{2}{\pi}\frac{\sigma_\perp^4}{(p_T^2 + \sigma_\perp^2)^3}\frac{1}{\sqrt{2\pi}\sigma_z} e^{-\frac{p_z^2}{2\sigma_\parallel^2}}. \quad (30)$$

We still take the transverse width to be $\sigma_\perp = 300$ MeV$/c$ and have $\langle \partial/\partial p_z^2 \rangle = -\frac{1}{2}\sigma_\parallel^{-2}$ while now $\langle \partial/\partial p_x^2 \rangle = \langle \partial/\partial p_y^2 \rangle = -2\sigma_\perp^{-2}$. [We have also considered replacing the longitudinal profile by a Gaussian in the rapidity $y = \frac{1}{2}\ln[(E+p_z)/(E-p_z)]$, but such a form is unsuitable because of its singular behavior at $p_\perp = 0$.]

Figure 1 illustrates shape of these two test profiles. Generally, for similar values of $\sigma_\perp$ and $\sigma_\parallel$, the pQCD form exhibits a slower transverse falloff which (as we shall see) enhances the degree of instability.

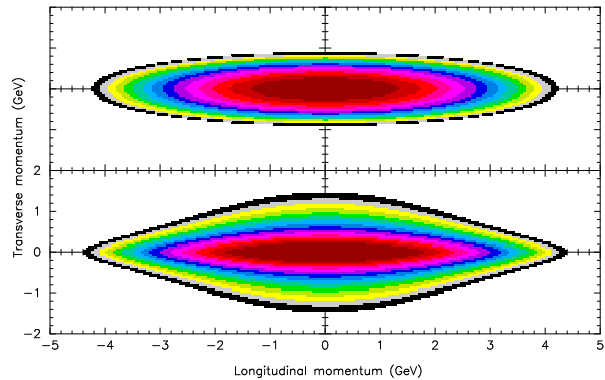

FIG. 1: Contour plot of the momentum profile function $\phi(\mathbf{p})$ for either the Gaussian (*top*) or the pQCD (*bottom*) parametrization, using $\sigma_\perp = 300$ MeV$/c$ and $\sigma_\parallel = 1$ GeV$/c$. The presentation is logarithmic with four bands per decade.

## V. TRANSVERSE MODES

In order to illustrate the typical appearance of the dispersion relation, we show in Fig. 2 the growth rate $\gamma_\mathbf{k}$ for purely transverse modes $\mathbf{k} = (k_\perp, 0, 0)$ as a function $k_\perp$, for a few selected values of the momentum anisotropy in matter corresponding to the initial RHIC scenario. The curve $\gamma(k_\perp)$ rises linearly from $k_\perp = 0$ with an initial slope that increases with the anisotropy (see Eq. (26)). The curve then exhibits a maximum $\gamma_0$ at the optimal wave number $k_\perp = k_0$, and then decreases roughly linearly towards zero as $k_\perp$ approaches $k_\perp = k_{\max}$, with a slope that decreases as the anisotropy increases. Although the two different momentum profiles yield qualitatively similar dispersion relations, the slower transverse falloff of the pQCD form generally increases the degree of instability. As a consequence, the maximum wave number $k_{\max}$ becomes significantly larger and, perhaps more importantly, the maximum growth rate $\gamma_0$ is also increased, although this effect is only rather modest.

In order to provide a more global impression of these features, we first consider the dependence of the maximum wave number $k_{\max}$ on the various input values (see Fig. 3). For the Gaussian profile we have

$$\text{Gauss: } k_{\max}^2 = g^2 \rho_{\text{eff}} \langle v_z p_z \rangle \left(\frac{1}{\sigma_x^2} - \frac{1}{\sigma_z^2}\right), \quad (31)$$

which shows that there are unstable modes as long as $\sigma_z > \sigma_x$. The limiting case, $k_{\max} = 0$, corresponds to isotropy, $\sigma_z^2 = \sigma_x^2$ $(=\frac{1}{2}(300 \text{ MeV})^2)$. For the pQCD profile we find

$$\text{pQCD: } k_{\max}^2 = g^2 \rho_{\text{eff}} \langle v_z p_z \rangle \left(\frac{2}{\sigma_x^2} - \frac{1}{\sigma_z^2}\right). \quad (32)$$

Thus, evidently, $k_{\max}$ is now larger and the limiting case, $k_{\max} = 0$, is oblate, $\sigma_z^2 = \frac{1}{2}\sigma_x^2$ $(=\frac{1}{4}(300 \text{ MeV})^2)$. The resulting values of $k_{\max}$ are shown in Fig. 3. Since $k_{\max}^2 \sim \alpha_s \rho_{\text{eff}}$, the results for the LHC and RHIC scenarios differ by a factor of $[(\alpha\rho)_{\text{LHC}}/(\alpha\rho)_{\text{RHIC}}]^{1/2} = \frac{5}{3}$, in reflection of the general scaling properties of the dispersion equation (24). Furthermore, since $\langle v_z p_z \rangle \approx \langle p_z \rangle = \sqrt{2/\pi}\sigma_z$, we obtain the following fairly accurate approximation,

$$k_{\max}^2 \approx g^2 \rho_{\text{eff}} \sqrt{\frac{2}{\pi}} \sigma_z \left( \frac{1 \text{ or } 2}{\sigma_x^2} - \frac{1}{\sigma_z^2} \right) \sim \sigma_z . \quad (33)$$

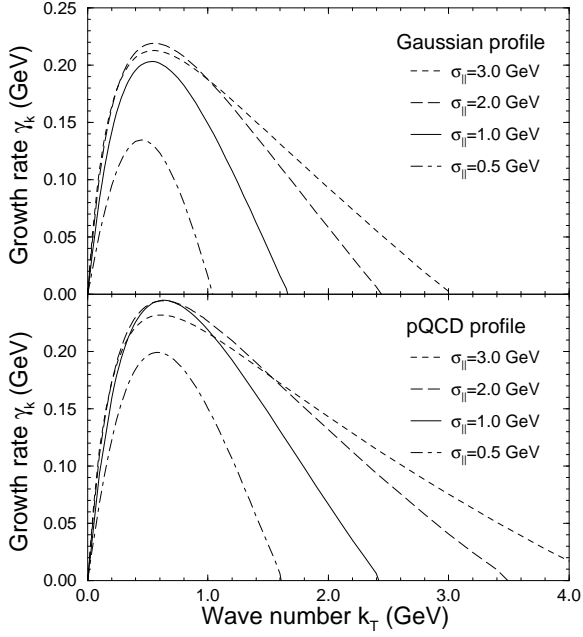

FIG. 2: The dispersion relation for four initial RHIC scenarios ($\alpha_s=0.3$, $\rho_{\text{eff}}=6$ fm$^{-3}$) with various momentum anisotropies: the growth rate $\gamma_{\mathbf{k}}$ for purely transverse modes, $\mathbf{k}=(k_\perp,0,0)$, as a function of the transverse wave number $k_\perp$, for the Gaussian (*top*) or pQCD (*bottom*) momentum profile $\phi(\boldsymbol{p})$.

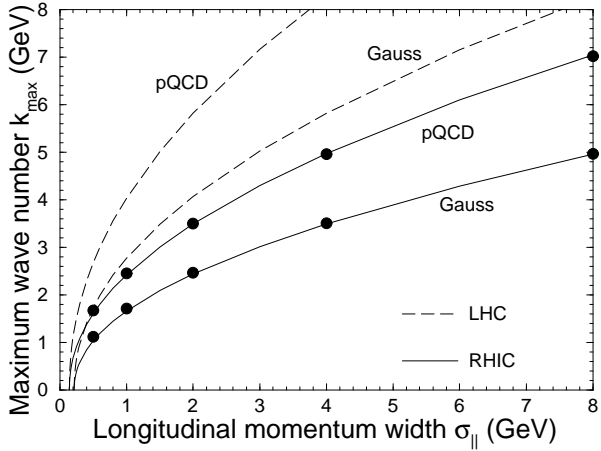

FIG. 3: The maximum wave number $k_{\max}$ at the initial time $t_0$ as a function of the longitudinal momentum dispersion $\sigma_\parallel$ for Gaussian and pQCD momentum profiles $\phi(\boldsymbol{p})$ in the standard RHIC and LHC scenarios given in Eqs. (27-28). The approximate values given by Eq. (33) are shown for the RHIC scenario (dots).

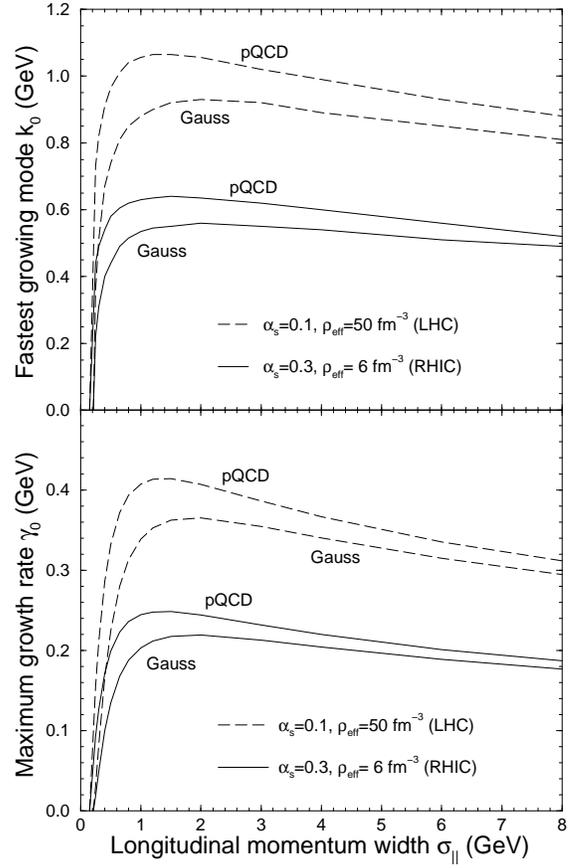

FIG. 4: The wave number of the fastest-growing mode, $k_0$ (*top*), and the associated maximum growth rate, $\gamma_0$ (*bottom*), as functions of the longitudinal momentum dispersion $\sigma_\parallel$ for RHIC (solid) and LHC (dashed) scenarios, unsing either Gaussian or pQCD momentum profiles having $\sigma_\perp = 0.3$ GeV/c.

More important than $k_{\max}$ is $k_0$, the wave number of the fastest-growing modes, since these modes will become predominant in the course of time. It is shown in Fig. 4 (*top*). After a sharp initial rise, $k_0$ exhibits a very slow decrease with increasing anisotropy. For a wide range of $\sigma_\parallel$ values, the wave number of the fastest growing modes obtained with the Gaussian profile is typically around 500 MeV at RHIC and 900 MeV at LHC, corresponding to wave lengths $\lambda_0$ of 2.5 fm and 1.4 fm, respectively. For the pQCD profile these wave numbers are typically 100 MeV higher. We note that unless the anisotropy is small, we have $k_0 \ll k_{\max}$.

The behavior of the maximum growth rate $\gamma_0$ is shown in Fig. 4 (*bottom*) and it is qualitatively similar to that of $k_0$: it exhibits a rapid initial growth followed by a slow decrease. The maximum growth rate is obtained for $\sigma_\parallel \approx 1.5 - 2.0$ GeV/c and the corresponding overall shortest growth times $\tau_0 = 1/\gamma_0$ are 0.90 fm/c and 0.55 fm/c for the Gaussian profile, respectively, while they are about 20% shorter for the pQCD profile. Thus the pQCD profile, with its slower transverse falloff, leads to somewhat larger growth rates. However, the overall results are very similar for the two profile forms. In Fig. 4 as well, the scaling feature of the dispersion equation (24) implies that the LHC curves are simply a factor of $\frac{5}{3}$ larger than the RHIC curves. It is thus fairly easy to infer the behavior of the dispersion relation for any other particular choices of the input parameters $\alpha_s$ and $\rho_{\text{eff}}$.

Our results for the transverse modes agree qualitatively with the analytical results obtained in Ref. [11].

## VI. GENERAL MODES

As noted above, for any given purely perpendicular wave vector $\mathbf{k}=(k_\perp,0,0)$ for which there exists a filamentation mode (whose induced current is then perfectly aligned with the symmetry axis), there exists an entire range of misaligned modes that are also unstable, being characterized by the wave vector $\mathbf{k}=(k_\perp,0,k_\parallel)$, where $k_\parallel < k_\parallel^{\max}(k_\perp)$.

The dispersion equation for these general filamentation modes is given by Eq. (22). Since $D_\mathbf{k}(\gamma)$ is even and vanishes for $\gamma=0$, there is at most one positive root for a given $\mathbf{k}$. The resulting maximum values of $k_\parallel$ are shown in Fig. 5 for selected values of the anisotropy, in both the RHIC and LHC scenarios. These curves delineate the respective spinodal boundaries in $\mathbf{k}$ space and the isotropic metric employed in the display ensures that the directional information is meaningful.

For a general mode, whose wave vector $\mathbf{k}$ is not perpendicular to the $z$ axis, the situation is more complicated and the associated fields and currents are no longer simply related to $\mathbf{k}$. In order to illutrate the increased complexity, we consider here the $\mathbf{k}$ dependence of the electric field strength $\boldsymbol{E}$ and the color current density $\boldsymbol{j}$ in the Coulomb gauge. (The results are gauge invariant for electromagnetic plasmas.) Thus, the electric field $\boldsymbol{E}$ is determined by the following $3\times 3$ matrix equation,

$$[\mathbf{k}^2\delta^{ij} - k^i k^j - \omega^2 \epsilon^{ij}(k)]E_a^j(k) = 0 . \quad (34)$$

Since the $yy$ term never vanishes (and the off-diagonal terms involving $y$ vanish by symmetry), we must have $E_y=0$. Thus the field vector lies in the $xz$ plane, the plane spanned by the wave vector and the symmetry axis of the momentum distribution $f(\boldsymbol{p})$. In the aligned case, the wave vector is perpendicular to the symmetry axis, $\mathbf{k}=(k_x,0,0)$, and the resulting electric field is directed along the symmetry axis, $\boldsymbol{E}=(0,0,E_z)$. In the general case, the wave vector has a component parallel to the symmetry axis, $\mathbf{k}=(k_x,0,k_z)$, and thus no longer lies in the transverse plane. The angle formed with the symmetry axis, $\vartheta_\mathbf{k}$, is determined by $\tan\vartheta_\mathbf{k}=k_x/k_z$, while the polar angle of the associated electric field vector, $\vartheta_\mathrm{E}$, is determined by

$$\tan\vartheta_\mathrm{E} = \frac{E_x}{E_z} = \frac{k_x k_z - \Pi_{xz}}{k_z^2 + \gamma^2 + \Pi_{xx}} = \frac{k_x^2 + \gamma^2 + \Pi_{zz}}{k_z k_x - \Pi_{zx}} . \quad (35)$$

For small values of $k_z$, $\vartheta_\mathbf{k}$ is near $90°$, while $\vartheta_\mathrm{E}$ is near $0°$. These directions evolve steadily as the magnitude of $k_z$ is increased, $\vartheta_\mathbf{k}$ decreasing and $\vartheta_\mathrm{E}$ increasing, until they become equal just as the maximum value of $k_z$ is reached, $\vartheta_\mathbf{k}(k_\perp,k_\parallel^{\max}(k_\perp)) = \vartheta_\mathrm{E}(k_\perp,k_\parallel^{\max}(k_\perp))$. It follows that $\boldsymbol{E}\perp\mathbf{k}$ *only* when $\mathbf{k}$ is perpendicular to the symmetry axis and thus it is only those modes that have a purely transverse character. Furthermore, there are no purely longitudinal modes (for which $\boldsymbol{E}\parallel\mathbf{k}$).

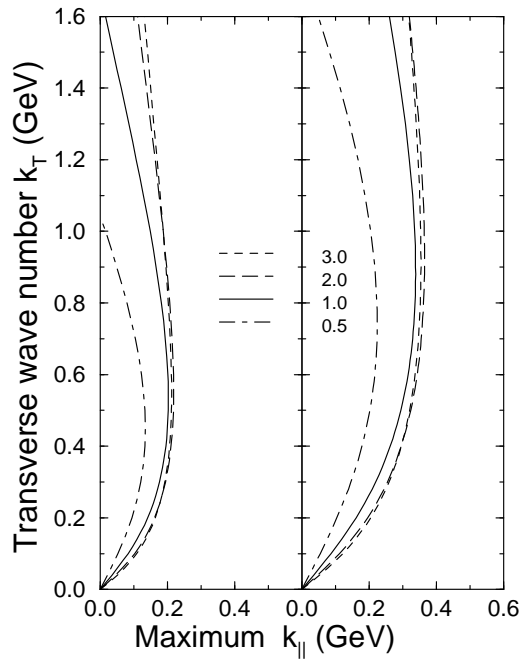

FIG. 5: The maximum value of the longitudinal wave number, $k_\parallel^{\max}$, as a function of the magnitude of the transverse wave number, $k_\perp$ (a mode $\mathbf{k}=(k_\perp,0,k_\parallel)$ is unstable if $k_\parallel < k_\parallel^{\max}$), for the initial RHIC (*bottom*) and LHC (*top*) scenarios. The momentum profile is Gaussian and the various values of the longitudinal momentum dispersion $\sigma_z$ are indicated. The vertical and horizontal scales are equal.

For large values of $k_x$, where $\gamma$ becomes small, there is not much room for $k_z$ and the limiting direction does not deviate much from the $x$ direction (it approaches the $x$ direction when $k_x$ approaches $k_{\max}$). Obviously, there is most room for $k_z$ for those $k_x$ values that are in the region of maximum growth, while for small $k_x$, where also $\gamma$ tends to zero, there is again little room for $k_z$. However, since $k_x$ is now also small, the direction of $\mathbf{k}$ is significantly affected by the addition of $k_z$. The resulting unstable region in $\mathbf{k}$ space is thus a flat disk-like volume (with a central depression) which is oriented perpendicular to the symmetry axis of the momentum profile (the beam axis).

It also follows that $\boldsymbol{\Pi}\cdot\mathbf{k}=0$ at the spinodal boundary. This feature implies that the limiting direction of the current is perpendicular to the wave vector (and hence to the electric field as well). Furthermore, the direction of the current $\boldsymbol{j}$ turns in the same sense as the wave vector $\mathbf{k}$, but initially at a slower rate, so their relative angle never exceeds 90°, $\mathbf{k} \cdot \mathbf{j} \geq 0$.

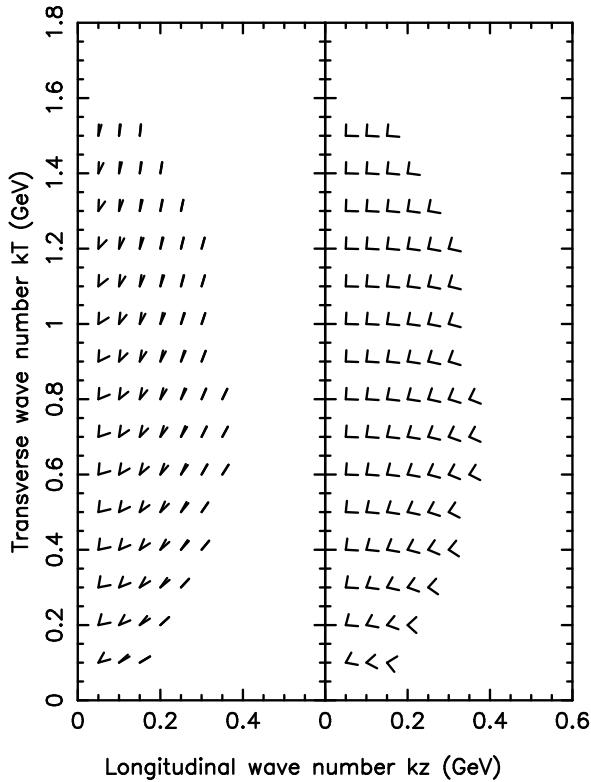

FIG. 6: For a lattice of wave vectors $\mathbf{k} = (k_\perp, 0, k_\parallel)$ are shown the direction of the wave vector, $\mathbf{e}_k$ (the upper-left line of any pair), together with the direction of either the electric field, $\mathbf{e}_E$ (*left panel*), or the current density $\mathbf{e}_j$ (*right panel*), in the standard RHIC scenario with a Gaussian momentum profile having $\sigma_\parallel = 1$ GeV/$c$. These results have been obtained in the Coulomb gauge.

The evolution of these various directions with the wave vector $\mathbf{k}$ is illustrated in Fig. 6. We see that in the region of largest amplification the degree of reorientation can be substantial, typically amounting to $\Delta\vartheta \approx 30°$. The resulting modes are rather complicated to describe, since the directions of the $\mathbf{k}$, $\mathbf{E}$, and $\mathbf{j}$ have no simple mutual relationship.

## VII. AMPLIFICATION COEFFICIENTS

The environments produced in high-energy nuclear collisions are endowed with a rapid expansion, primarily in the longitudinal (beam) direction $\hat{z}$. The effective density then decreases towards zero in the course of time, $\rho_{\text{eff}}(t) \to 0$. We shall therefore solve the dispersion equation at successive times $t \geq t_0$ and thus obtain a time-dependent growth rate $\gamma_\mathbf{k}(t)$ for each wave vector $\mathbf{k}$. (Such an adiabatic approach is valid only for sufficiently slow evolutions and the results should therefore be regarded as approximative.) Since the maximum wave number for which instabilities exist, $k_{\max}$, is proportional to the square root of $\rho_{\text{eff}}$, it also decreases in time, $k_{\max}(t)$. Consequently, for each particular wave vector $\mathbf{k}$, there is a time, $t_\mathbf{k}^{\max}$ beyond which the associated collective frequencies $\pm\omega_\mathbf{k}$ are real.

Within the adiabatic approximation, the amplitude of a given unstable mode, $C_\mathbf{k}^\pm$, evolves as

$$C_\mathbf{k}^\pm(t) = C_\mathbf{k}^\pm(t_0) \, \exp[\pm \int_{t_0}^t \gamma_\mathbf{k}(t')dt'] \,. \qquad (36)$$

The accumulated increase of the corresponding strength is then governed by the *amplification coefficient* [18, 19],

$$\Gamma_\mathbf{k} \equiv \int_{t_0}^{t_\mathbf{k}^{\max}} \gamma_\mathbf{k}(t) \, dt \,. \qquad (37)$$

An elementary analysis shows that if the density falls off as an inverse power of time, $\rho_{\text{eff}} \sim t^{-\beta}$, then $\Gamma_\mathbf{k} \sim k^{1-2/\beta}$ in the limit of soft modes. Thus the amplification coefficient diverges unless the falloff is at least quadratic.

In a high-energy nuclear collision, the falloff of the density is initially approximately inversely proportional to time, $\beta \approx 1$, due to the rapid longitudinal expansion, but it then quickens (ultimately to $\beta \approx 3$) as the transverse expansion manifests itself. Thus, if carried through, the adiabatic treatment would yield finite amplification coefficients for all modes $\mathbf{k}$. For simplicity, and to avoid a sensitive dependence on the long-time behavior for the soft modes, we shall employ a simple exponential falloff,

$$\rho_{\text{eff}}(t) \approx \rho_{\text{eff}}(t_0) \, e^{-\delta t/t_0} = \rho_{\text{eff}}(t_0) \, [\frac{t_0}{t} + \mathcal{O}(\frac{t_0^2}{t^2})] \,, \quad (38)$$

where $\delta t \equiv t - t_0$ is the elapsed time. This form matches the initial longitudinal expansion while ensuring that $\gamma_\mathbf{k}(t)$ drops off sufficiently fast at large times to avoid soft divergencies. It thus allows us to evaluate the amplification coefficients in a manner that is insensitive to the long-term behavior of the collision system.

We first calculate $\Gamma_\mathbf{k}$ under the assumption that the momentum distribution remains constant in time while $\rho_{\text{eff}}$ drops off in the above exponential fashion (38). This is expected to provide an overestimate of the amplification, since both expansion and equilibration act to reduce the anisotropy. The resulting amplification coefficients are shown in Fig. 7. The curves all exhibit a steep initial rise followed by a gentle descent for large wave numbers. Since the modes with small wave numbers are subject to amplification for a longer time, the curves have their maxima shifted downwards relative to the $k_0$ values for the respective initial scenarios. Thus, the largest amplification coefficients are obtained for $k_\perp \approx 250$ GeV/$c$ in the RHIC scenario and for $k_\perp \approx 400$ GeV/$c$ in the LHC scenario. Their largest values are about 0.7 and 0.9, respectively, and they are reached for $\sigma_\parallel \approx 2 - 4$ GeV/$c$.

Since the pCDQ momentum profile generally leads to growths rates $\gamma_\mathbf{k}$ that are larger than those obtained with the Gaussian profile (see above), the resulting amplification coefficients $\Gamma_\mathbf{k}$ are correspondingly larger, by about 20% in the region of maximum amplification.

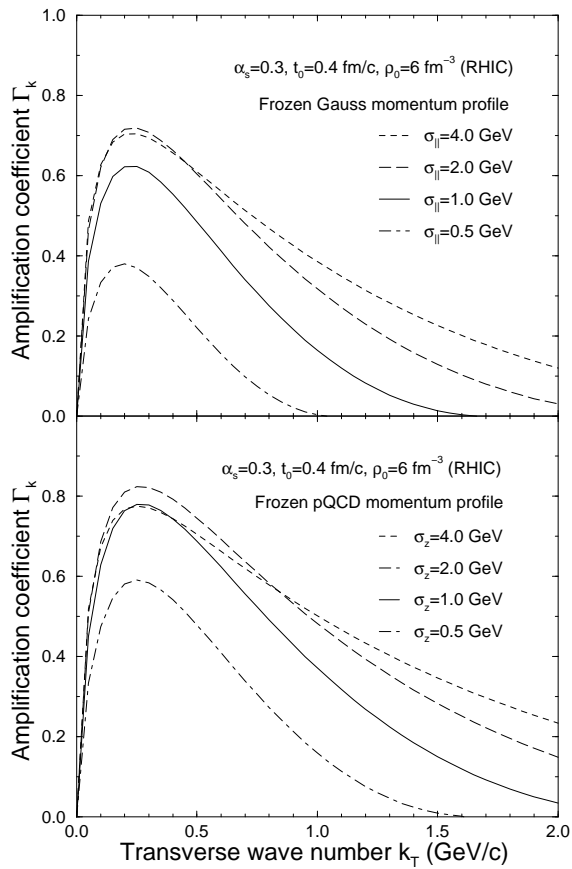

FIG. 7: The amplification coefficient $\Gamma_{\bm{k}}$ for purely transverse Weibel modes as a function of the wave number $k_\perp$, for the idealized case when the momentum profile remains frozen throughout while the density decreases exponentially according to Eq. (38); calculated for the RHIC scenario with either Gaussian (*top*) or pQCD (*bottom*) profiles having the specified longitudinal widths $\sigma_z$.

## VIII. MOMENTUM RELAXATION

In the above analysis, we have taken account of the decreasing density but kept the momentum profiles constant. A more realistic treatment must take account of the dynamical evolution of the momentum profile as well. Of particular importance are the rapid longitudinal expansion and the possibility of collisions among the partons. In order to elucidate the quantitative importance of these agencies, we allow the momentum distribution to evolve as it would if subjected to a combination of idealized longitudinal expansion and elastic Boltzmann collisions, while still assuming that the overall effective density behaves as in (38). (Thus, although we have neglected the possible influence of the parton-parton collisions on the dispersion relation, we do now consider their effect on the anisotropy in the medium.)

If we ignore the effect of the expansion and treat the collisions in the relaxation-time approximation, the equation of motion for the momentum distribution is simple,

$$\frac{\partial}{\partial t}\phi(\bm{p},t) = -\frac{1}{t_c}[\phi(\bm{p},t) - \tilde{\phi}(p)] , \qquad (39)$$

where $\tilde{\phi}(p)$ is the equilibrium profile (which is isotropic) and $t_c$ is the relaxation time. If we are only interested in the momentum variances, $\sigma_i^2(t) \equiv \int d^3\bm{p}\, p_i^2 \phi(\bm{p},t)$, then Eq. (39) reduces to a set of coupled equations,

$$\frac{\partial}{\partial t}\sigma_i^2(t) = -\frac{1}{t_c}[\sigma_i^2(t) - \tilde{\sigma}^2] , \qquad (40)$$

where $\tilde{\sigma}^2$ is the equilibrium variance. The evolution is then the familiar exponential relaxation,

$$\sigma_i^2(t) = [\sigma_i^2(t) - \tilde{\sigma}^2]\,e^{-\frac{t-t_0}{t_c}} + \tilde{\sigma}^2 . \qquad (41)$$

We now include a longitudinal scaling expansion, which causes the density to decrease steadily in time, $\rho(t) = \rho(t_0)t_0/t$. [The scaling scenario is boost invariant, so it suffices to consider what happens in a rest frame at the origin, where $t$ and $z$ are identical to the general variables $\tau$ and $\eta$.] Since the collision rate is inversely proportional to the density, $t_c^{-1} = \rho\sigma\bar{v}$, it is reasonable to assume that it exhibits a similar decrease, $t_c^{-1} = C/t$. The equations of motion (40) are then modified,

$$\frac{\partial}{\partial t}\sigma_x^2 = -\frac{C}{t}[\sigma_x^2 - \tilde{\sigma}^2] , \qquad (42)$$

$$\frac{\partial}{\partial t}\sigma_z^2 = -\frac{C}{t}[\sigma_z^2 - \tilde{\sigma}^2] - \frac{1}{t}\sigma_z^2 , \qquad (43)$$

The last term in (43) seeks to flatten the local momentum distribution in response to the longitudinal stretching, which at the same time reduces the collision rate.

The resulting dynamics is then more complicated. For simplicity, we shall assume that the sum of the momentum variances is preserved by the Boltzmann collisions, as in elastic non-relativistic collisions. The equilibrium variance at a given time $t$ is then given by $\tilde{\sigma}^2 = \frac{1}{3}(2\sigma_x^2 + \sigma_z^2)$. The equations of motion (42-43) can then be rewritten on matrix form,

$$\frac{\partial}{\partial t}\begin{pmatrix}\sigma_x^2\\ \sigma_z^2\end{pmatrix} = -\frac{1}{3t}\begin{pmatrix} C & -C \\ -2C & 3+2C \end{pmatrix}\begin{pmatrix}\sigma_x^2\\ \sigma_z^2\end{pmatrix} . \qquad (44)$$

The eigenvalues of the time-independent coupling matrix, $\lambda_\mp$, are determined by the following secular equation,

$$\lambda_\mp^2 - 3(1+C)\lambda_\mp + 3C = 0 , \qquad (45)$$

and the eigenvalues of Eq. (44) are $\Lambda_\mp = -\lambda_\mp/3t$, i.e.

$$\Lambda_\mp(t) = \frac{1}{2t}\left\{1+C \mp \left[(1+C)^2 - \frac{4}{3}C\right]^{\frac{1}{2}}\right\} > 0 . \qquad (46)$$

It then follows that the associated normal variances, $\sigma_\mp^2$, fall off as inverse powers of time,

$$\frac{\sigma_\mp^2(t)}{\sigma_\mp^2(t_0)} = e^{\int_{t_0}^t \Lambda_\mp(t')dt'} = e^{-\frac{\lambda_\mp}{3}\int_{t_0}^t \frac{dt'}{t'}} = \left(\frac{t_0}{t}\right)^{\frac{\lambda_\mp}{3}} . \qquad (47)$$

They represent approximately the distortion $\sigma_z^2 - \sigma_x^2$ and (half) the total variance $2\sigma_x^2 + \sigma_z^2$, respectively.

Since the Cartesian variances can be expressed in terms of the normal variances,

$$\sigma_x^2 = \mathcal{N}[\sigma_-^2 - \epsilon\sigma_+^2] \ , \ \sigma_z^2 = \mathcal{N}[\sigma_+^2 + 2\epsilon\sigma_-^2] \ , \quad (48)$$

with $\mathcal{N}^{-1} = 1 + 2\epsilon^2$ and

$$\epsilon = \frac{1}{2C}\left\{\tfrac{3}{2}\left[1 + \tfrac{2}{3}C + C^2\right]^{\frac{1}{2}} - \tfrac{3}{2} - \tfrac{1}{2}C\right\} > 0 \ , \quad (49)$$

their time evolution can be readily obtained. It is seen that all the variances tend to zero for any (positive) value of the collision constant $C$, thereby making it possible to simultaneously achieve the continual longitudinal shrinkage caused by the expansion, $\sigma_z \to 0$, and the approach to isotropy caused by the collisions, $\sigma_z \to \sigma_x$.

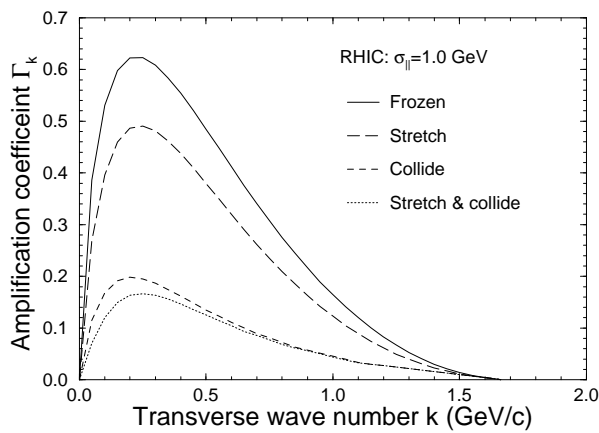

FIG. 8: The effect of Bjorken expansion and Boltzmann relaxation on the amplification coefficient $\Gamma_\mathbf{k}$, for the RHIC case with $\sigma_\parallel(t_0) = 1$ GeV/$c$ in various dynamical scenarios: Frozen momentum distributions, as in Fig. 7 (solid); flattening of the momentum distribution due to the stretching of the system (long dashes); relaxation of the momentum distribution due to the collisions, assuming $C = 1$ so $t_c(t_0) = 0.4$ fm/$c$ (short dashes); and both of those agencies active (dots).

The effects of the expansion and equilibration on the amplification coefficients is illustrated in Fig. 8 for Gaussian momentum profiles. In view of our previous findings, we expect that the effects are similar for pQCD profiles. We note that the stretching produces an only relatively moderate decrease (about 20%), whereas the collisions reduce the amplification coefficients by about a factor of three. The combined effect is then a reduction of $\Gamma_\mathbf{k}$ by a factor of nearly four. The collision rate has been somewhat arbitrarily set to $C = 1$, corresponding to an initial Boltzmann relaxation time of $t_c(t_0) = t_0$. Our results show that there is a considerable sensitivity to this quantity which might be better estimated from microscopic parton cascade models [20].

In order to further elucidate the effect of scaling expansion and Boltzmann relaxation, we show in Fig. 9 the effect of including both in the approximately optimal RHIC and LHC scenarios, for which the initial longitudinal variance is around $\sigma_\parallel = 2$ GeV/$c$ (see Fig. 7). As already noted in connection with Fig. 8, the inclusion of these effects lead to an overall reduction of more than a factor of four, relative to the simple idealized case in which the momentum profiles remain frozen as the density decreases according to Eq. (38) (shown in Fig. 7). The results obtained with pQCD profiles are quite similar, apart from the values being overall slightly larger (about 20% for the most amplified modes).

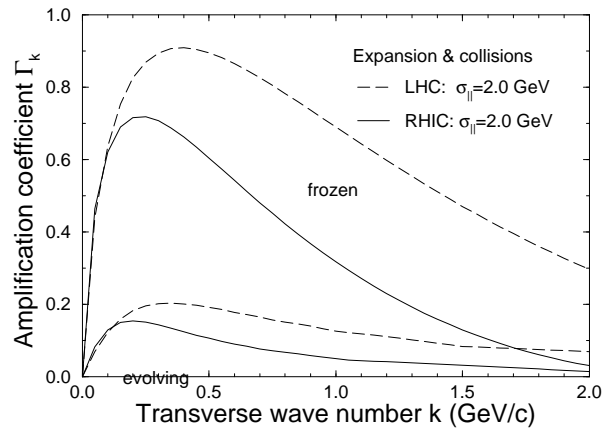

FIG. 9: The effect of scaling expansion and Boltzmann equilibration on the amplification coefficient $\Gamma_k$ for RHIC and LHC scenarios. The momentum profiles are Gaussian and have an initial dispersion of $\sigma_\parallel(t_0) = 2$ GeV/$c$, for which the largest net amplification is obtained. The solid curves are those obtained for frozen profiles (shown in Fig. 7), while the dashed curves are obtained in the presence of both scaling expansion and Boltzmann collisions (with $C = 1$).

Of course, if the Boltzmann relaxation time is increased, the effect of the collisions will decrease, and vice versa. Since the actual collision rate is hard to assess, it may only be safe to conclude that the collisions are quantitatively important, while the (somewhat more complicated) effect of the expansion appears to be less crucial.

## IX. CONCLUDING REMARKS

In the present study, we have sought to make quantitative estimates of the importance of Weibel instabilities in the chromodynamic plasma created early on in a high-energy nuclear collision. Any quantitative calculation must rely on specific assumptions about the plasma environment, including its dynamical evolution which is presently only rather poorly known. Therefore, our results are correspondingly approximate. Fortunately, though, the simple scaling properties of the Weibel dispersion relation makes it relatively easy to infer what the result would have been if different input values had been employed. Therefore, the utility of our results extends beyond the specific cases presented as illustrations.

10As a concrete framework for our discussion, we have adopted two standard scenarios, one appropriate for RHIC and the other for LHC, in terms of the coupling constant $\alpha_s$, the initial effective parton density $\rho_{\text{eff}}(t_0)$, and the corresponding starting time $t_0$. Furthermore, we have employed axially symmetric momentum profiles that are either Gaussian or pQCD-motivated (leading to a power falloff in the transverse direction). We have then kept the transverse momentum dispersion constant at $\sigma_\perp = 300$ MeV/$c$, while exploring the dependence on the (local) longitudinal spread $\sigma_\parallel$ (for a given density and profile type, the results depend only on the ratio $\sigma_\parallel/\sigma_\perp$).

Going beyond earlier treatments of these modes, we have permitted the wave vector $\mathbf{k}$ to have a component along the symmetry axis as well, thus extending the considerations to modes that are not purely transverse. The associated polarization tensor is then no longer diagonal. Generally the growth rates, $\gamma_\mathbf{k}$, decrease as the parallel component of $\mathbf{k}$ is increased and the resulting region of instability in $\mathbf{k}$ space is widest in the region where the aligned modes have their maximal growth rates. Furthermore, the electric field of a given mode, $\boldsymbol{E}_\mathbf{k}$, as well as the induced current, $\boldsymbol{j}_\mathbf{k}$, form an angle with the symmetry axis, as we illustrated in the Coulomb gauge. The electric field then turns *towards* $\mathbf{k}$ and finally becomes parallel to $\mathbf{k}$ at the spinodal boundary, while the current density turns in the same sense as $\mathbf{k}$. Being initially aligned with the symmetry axis (and thus perpendicular to $\mathbf{k}$), it turns at first at a slightly slower rate, so its angle with $\mathbf{k}$ becomes smaller than 90°. It then starts turning at a faster rate and reverts to being perpendicular to $\mathbf{k}$ at the boundary, where it is thus perpendicular to $\boldsymbol{E}$.

The largest growth rates $\gamma_\mathbf{k}$ are obtained for modes whose wave numbers are perpendicular to the symmetry axis and they have a transverse character, $\boldsymbol{E} \perp \mathbf{k}$. Over a wide range of $\sigma_\parallel$ values, these wave numbers are typically around 500 MeV at RHIC and 900 MeV at LHC, corresponding to wave lengths $\lambda_0$ of 2.5 fm and 1.4 fm, respectively. The corresponding growth rates are generally larger for the pQCD profiles, but only by about 20% for the fastest modes. However, as the density decreases in the course of time, the higher wave numbers are progressively disfavored, as the spinodal region contracts. Therefore, the resulting amplification coefficient, $\Gamma_\mathbf{k} = \int \gamma_\mathbf{k}(t) dt$, which governs the accumulated degree of collective growth for a given $\mathbf{k}$, peaks at somewhat lower wave numbers, namely at $k \approx 250$ GeV/$c$ in the RHIC scenario and at $k \approx 400$ GeV/$c$ in the LHC scenario. In the idealized case when the momentum profiles $\phi(\boldsymbol{p})$ are kept frozen in time, the largest values of $\Gamma_\mathbf{k}$ are reached for $\sigma_\parallel \approx 2 - 4$ GeV/$c$ and amount to about 0.7 and 0.9, respectively, for Gaussian profiles and about 20% more for pQCD profiles. The inclusion of a longitudinal scaling expansion reduces these numbers only moderately by (by about 20%), so this dynamical complication is not so crucial. By contrast, the inclusion of elastic Boltzmann collisions among the partons leads to a significant reduction in the degree of instability, as the associated relaxation drives the momentum profile towards isotropy. For the adopted schematic collision rate, which corresponds to an initial relaxation time of $t_c(t_0) = t_0$, the reduction amounts to roughly a factor of three.

Thus, overall, we find that the degree of amplification of the Weibel filamentation modes is not expected to be spectacular for any particular wave vector $\mathbf{k}$. On the other hand, it appears that the effect may not be negligible either. Furthermore, it should be kept in mind that there are typically a large number of such unstable collective modes, so their combined effect on the overall dynamics may be significant.

We therefore wish to conclude by speculating about the possible dynamical consequences of the color filamentation phenomenon. One obvious aspect concerns the energy dissipation. Since the agitation of these collective modes would drain energy from the background system, the occurrence of color filamentation presents an additional agency for energy dissipation. Therefore, in principle, to the extent that these modes are agitated, one may expect a correspondingly faster equilibration of the parton system.

Furthermore, since the perfect azimuthal symmetry in an idealized head-on collision will be spontaneously broken by the appearance of the color currents, one may generally expect that the emergent filamentation pattern will manifest itself in the angular correlations among the final hadrons. In particular, a non-statistical distribution of collective energy flow will emerge along the local Poynting vectors associated with each amplified filamentation mode. This expectation is qualitatively different from that based on the parton cascade simulations [20]. The breaking of the azimuthal symmetry is then caused by jets produced in hard parton-parton interactions and, consequently, the effect is carried by only a few partons with large transverse momenta. By contrast, due to the collective character of the filamentation instability, the azimuthal symmetry breaking will be presumably involve a large number of partons having relatively small transverse momenta.

It has already been speculated that color filamentation may have observables consequences for the elliptic flow [21], based on the argument that the parton trajectories tend to become concentrated within the centers of the filaments. The conservation of phase-space volume then expands the momentum distribution perpendicular to the filaments. A corresponding quantum-mechanical argument can be made on the basis of the uncertainty relation.

Finally, it would appear that color filamentation might delay hadronization. This possibility is due to the basic fact that no hadronization can occur in the presence of color currents, since the hadronic phase is constituted entirely of colorless entities and thus unable to sustain any colored agitations. Thus, any collective color currents induced by the filamentation phenomenon would have to subside before the chromodynamic plasma could transform itself into an assembly of color singlets. It would thus be of interest, in a future study, to estimate how quickly the induced color currents dissolve again.

In conclusion, while it appears that color filamentation may occur in the early parton plasma, a quantitative assessment of the significance of the phenomenon will require detailed dynamical treatments that are not yet sufficiently developed. As a step in this direction, it might be interesting to solve the self-consistent Vlasov equations (10-12) in schematic collision scenarios in order to investigate how the filamentation modes manifest themselves.

**Acknowledgements**

J.R. was supported by the Director, Office of Energy Research, Office of High Energy and Nuclear Physics, Nuclear Physics Division of the U.S. Department of Energy under Contract No. DE-AC03-76SF00098 and by the Kavli Institute of Theoretical Physics, University of California at Santa Barbara. St.M. was partially supported by the Polish Committee of Scientific Research under grant 2P03B04123.